\documentclass[10pt,twocolumn,letterpaper]{article}

\usepackage[pagenumbers]{cvpr} % To force page numbers, e.g. for an arXiv version

\usepackage{graphicx}
\usepackage{amsmath}
\usepackage{amssymb}
\usepackage{booktabs}
\usepackage{multirow}
\usepackage{overpic}

\usepackage[pagebackref,breaklinks,colorlinks]{hyperref}

\usepackage[capitalize]{cleveref}
\crefname{section}{Sec.}{Secs.}
\Crefname{section}{Section}{Sections}
\Crefname{table}{Table}{Tables}
\crefname{table}{Tab.}{Tabs.}

\usepackage{makecell}

\def\cvprPaperID{206} % *** Enter the CVPR Paper ID here
\def\confName{CVPR}
\def\confYear{2023}

\begin{document}

\title{SNAF: Sparse-view CBCT Reconstruction with Neural Attenuation Fields}

\author{
Yu Fang$^{1}$ \quad Lanzhuju Mei$^{1}$ \quad Changjian Li$^{2}$ \quad Yuan Liu$^{3}$ \\
Wenping Wang$^{4}$ \quad Zhiming Cui$^{1}$ \quad Dinggang Shen$^{1}$ 
\\
$^{1}$ShanghaiTech University \quad $^{2}$The University of Edinburgh \\
$^{3}$The University of Hong Kong \quad $^{4}$Texas A\&M University
}

\maketitle

%%%%%%%%% ABSTRACT
\begin{abstract}
Cone beam computed tomography (CBCT) has been widely used in clinical practice, especially in dental clinics, while the radiation dose of X-rays when capturing has been a long concern in CBCT imaging. Several research works have been proposed to reconstruct high-quality CBCT images from sparse-view 2D projections, but the current state-of-the-arts suffer from artifacts and the lack of fine details. In this paper, we propose {\em SNAF} for sparse-view CBCT reconstruction by learning the neural attenuation fields, where we have invented a novel view augmentation strategy to overcome the challenges introduced by insufficient data from sparse input views. Our approach achieves superior performance in terms of high reconstruction quality (i.e., {\em 30+} PSNR) with only $\mathbf{20}$ input views ({\em 25} times fewer than clinical collections), which outperforms the state-of-the-arts. We have further conducted comprehensive experiments and ablation analysis to validate the effectiveness of our approach.
\end{abstract}

%%%%%%%%% BODY TEXT
\section{Introduction}
\label{sec:intro}
\begin{figure}[!t]
  \centering
   \includegraphics[width=1.0\linewidth]{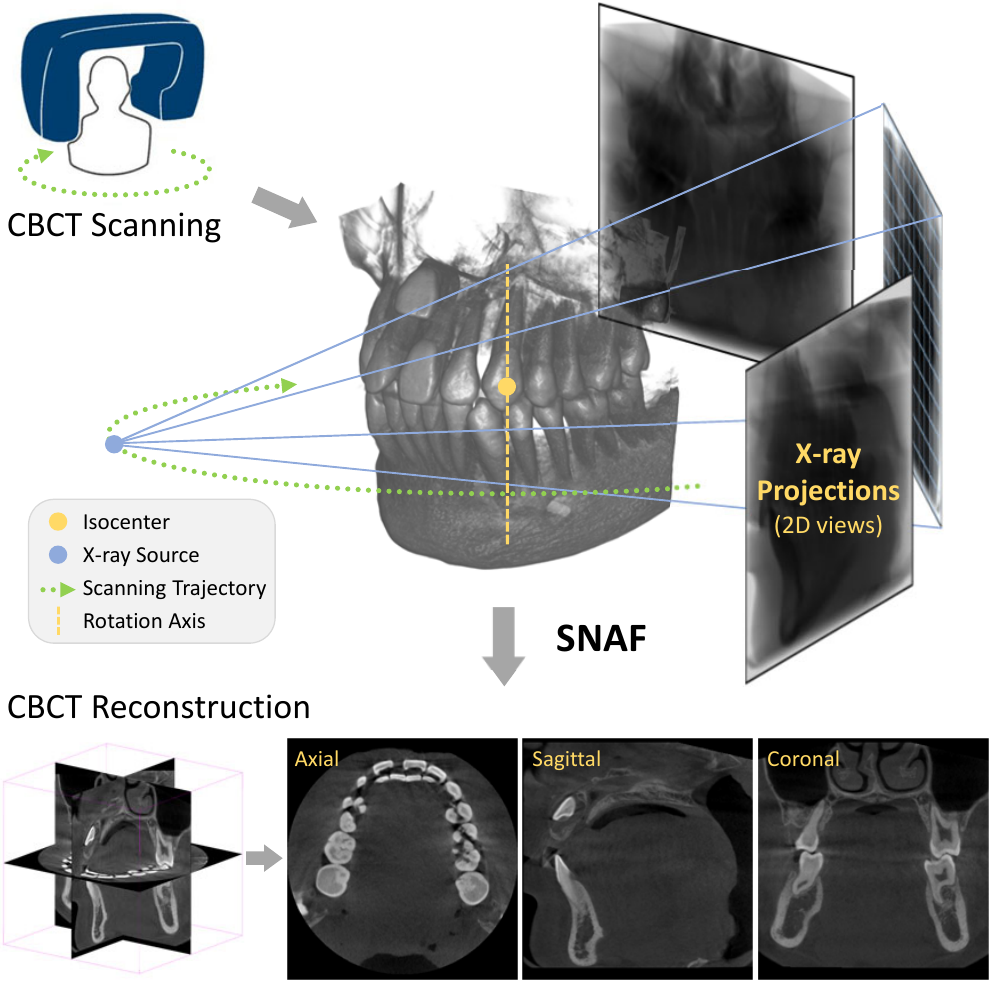}
    \caption{\textbf{Overview of sparse-view CBCT reconstruction with neural attenuation fields}. An X-ray source uniformly moves around the region of interest, and generates sparse projections with a flat-panel detector on the opposite side (top), while we propose {\em SNAF} to reconstruct the accurate CBCT from those sparse 2D views by learning the attenuation fields (bottom). }
   \label{fig:teaser}
\end{figure}

Cone beam computed tomography (CBCT) has become a major imaging technique in dental clinics (see Fig.~\ref{fig:teaser}).
Compared with conventional CT, it provides 3D information with higher quality and shorter scanning time~\cite{scarfe2006clinical}.
Those advantages, especially the high reconstruction quality, not only make it widely used for diagnosis and treatment planning in clinical practice, but also significantly facilitate research progresses for automatic CBCT image analysis (e.g., tooth and alveolar bone segmentation~\cite{cui2019toothnet, chung2020pose, cui2022fully}). 
However, as hundreds of views are required to be captured to ensure imaging quality, the radiation dose of X-rays has been a long concern in CBCT imaging.
To decrease the number of projections while maintaining the high imaging quality, several traditional methods for sparse-view reconstruction have been proposed - the typical filtered backprojection (FBP)~\cite{feldkamp1984practical}, the iterative methods~\cite{sidky2006accurate, cai2014cine} based on SART~\cite{andersen1984simultaneous}. These methods can produce decent reconstructed CBCT images, but still remain limitations such as the streaking artifacts and the lack of fine-details introduced by insufficient data.

Recently, neural rendering emerges to be an efficient technique for novel view synthesis and 3D scene reconstruction~\cite{mildenhall2020nerf}, benefiting from the implicit representation with compact structures.
Analogic to CBCT reconstruction, these methods utilize posed RGB images to learn underlying neural radiance fields (NeRFs), showing great advantages in 3D geometry awareness and multi-view consistency.
Follow-up works further improve their abilities for fast~\cite{liu2020neural, martel2021acorn, mueller2022instant} and sparse-view~\cite{yu2021pixelnerf, jain2021putting, niemeyer2022regnerf, kangle2021dsnerf} reconstruction.
These properties are highly expected to benefit sparse-view CBCT reconstruction.
However, unlike NeRFs that use volume rendering to approximate surface rendering, CBCT reconstruction goes back to authentic volume rendering. In this unique setting, there are no specific object surfaces as in the natural scene, and the entire space of human organs is informative, which significantly enlarges the solution space making the problem complex to be solved. Also, the popular regularization and generalization techniques introduced in sparse-view NeRFs, e.g., utilizing the additional depth information~\cite{kangle2021dsnerf}, are not applicable.

Inspired by our observations, we propose a novel framework called {\em SNAF} for sparse-view CBCT reconstruction by learning the attenuation fields, which achieves superior performance in terms of high reconstruction quality (i.e., {\em 30+} PSNR) with only $\mathbf{20}$ input views ({\em 25} times fewer than common clinical collections). Technically, our approach builds upon the state-of-the-art NeRF framework with a multi-resolution hash table~\cite{mueller2022instant}. To extend this framework for CBCT reconstruction, we remove the original importance sampling component, and design a ray casting strategy by equal-step marching to better integrate the information in the whole space. Most importantly, to deal with the main challenge of insufficient data from sparse input views, we propose a novel view augmentation strategy, which is based on our observation that the neighboring views contain rich shape and appearance information, providing strong regularization to constrain in-between novel views to be sharper and more clear. Specifically, we design a two-stage pipeline. In stage one, we first train a neural attenuation field network to render novel view images on predefined in-between viewpoints.
However, these novel view images are less trustable since they are usually blurry with little fine-details. We thus design a deblurring network to deblur these novel view images with the help of neighboring input views. Then, in stage two, the original input views along with the deblurred novel view images serve as inputs to further finetune the trained neural attenuation field. 

Our method achieves superior performance in terms of image quality, and we have conducted extensive experiments to demonstrate the effectiveness of our proposed SNAF. Additionally, we provide comprehensive ablation analysis to validate its stable performance under different scanning conditions, and to showcase its superior results on challenging data with metal artifacts and in real-world clinical applications.

\section{Related Work}
\subsection{Sparse-view Neural Radiance Fields}
\label{sec:rw_sparseview}
Neural rendering has achieved great performance in novel view synthesis and 3D scene reconstruction.
To extend NeRF~\cite{mildenhall2020nerf} in various aspects, inefficiency and requirements of dense views are two major shortcomings.
Many works accelerate the rendering process with more efficient volumetric representations, including sparse voxel grids~\cite{liu2020neural}, octree~\cite{martel2021acorn}, and hash table~\cite{mueller2022instant}.

Meanwhile, some approaches~\cite{chibane2021stereo,zhao2022humannerf,yu2021pixelnerf} are designed to consider sparse input scenarios.
For example, PixelNeRF~\cite{yu2021pixelnerf} and SRF~\cite{chibane2021stereo} extract CNN features from input views, so as to have the generalized ability to generate a novel scene with sparse inputs.
Concurrently, depth supervision has greatly improved reconstruction accuracy by regularizing geometry and appearance on novel views~\cite{kangle2021dsnerf, niemeyer2022regnerf}.
Unfortunately, they are not applicable to our CBCT reconstruction with neural attenuation fields.
First, generalization-based approaches are commonly designed to address cross-scene domain changes without inference optimization, where many 3D reconstructed details are usually lost which is unacceptable for clinical use.
Similarly, depth-based regularization is also not practical in our task, as there is no available depth or surface information under our transparent scenes.
In this paper, we propose a novel view augmentation strategy to confront sparse-view inputs, which is specifically devised under attenuation fields with their fixed patterns of input poses.

\subsection{CBCT Reconstruction}
High-quality CBCT reconstruction is crucial for clinical use and downstream computer-aided tasks, e.g., tooth and alveolar bone segmentation~\cite{cui2019toothnet, chung2020pose, cui2022fully}.
Filtered backprojection (FBP)~\cite{feldkamp1984practical}, most commonly used in commercial CBCT systems, is a traditional approach to accumulate the intensities by backprojection from given 2D views, yet requires hundreds of input views to prevent streaking artifacts.
This leads to high radiation dose and long scanning time, motivating the research and industry community to explore sparse-view reconstruction~\cite{andersen1984simultaneous, wang2004ordered}, for example, utilizing regularization to constraint the reconstruction problem.

Recently, as neural rendering has been rapidly developed, this emerging technique has been introduced to reconstruct CBCT images~\cite{ruckert2022neat, zha2022naf}.
These two works are concurrently proposed, adopting volumetric rendering for fast CBCT reconstruction.
However, NeAT~\cite{ruckert2022neat} has not been validated on clinical CBCT data scanned on human bodies, and NAF~\cite{zha2022naf} has not well explored the strategy for high-quality image reconstruction.
Moreover, these two methods do not address the steep challenging problem of sparse-view inputs.
In this work, we successfully address this problem with only 20 input views, and provide important discussions on different scanning conditions (e.g., number of input views, limited angles, and metal artifacts), as well as verification for downstream tasks (e.g., segmentation).

\begin{figure*}[t]
  \centering
   \includegraphics[width=1.0\linewidth]{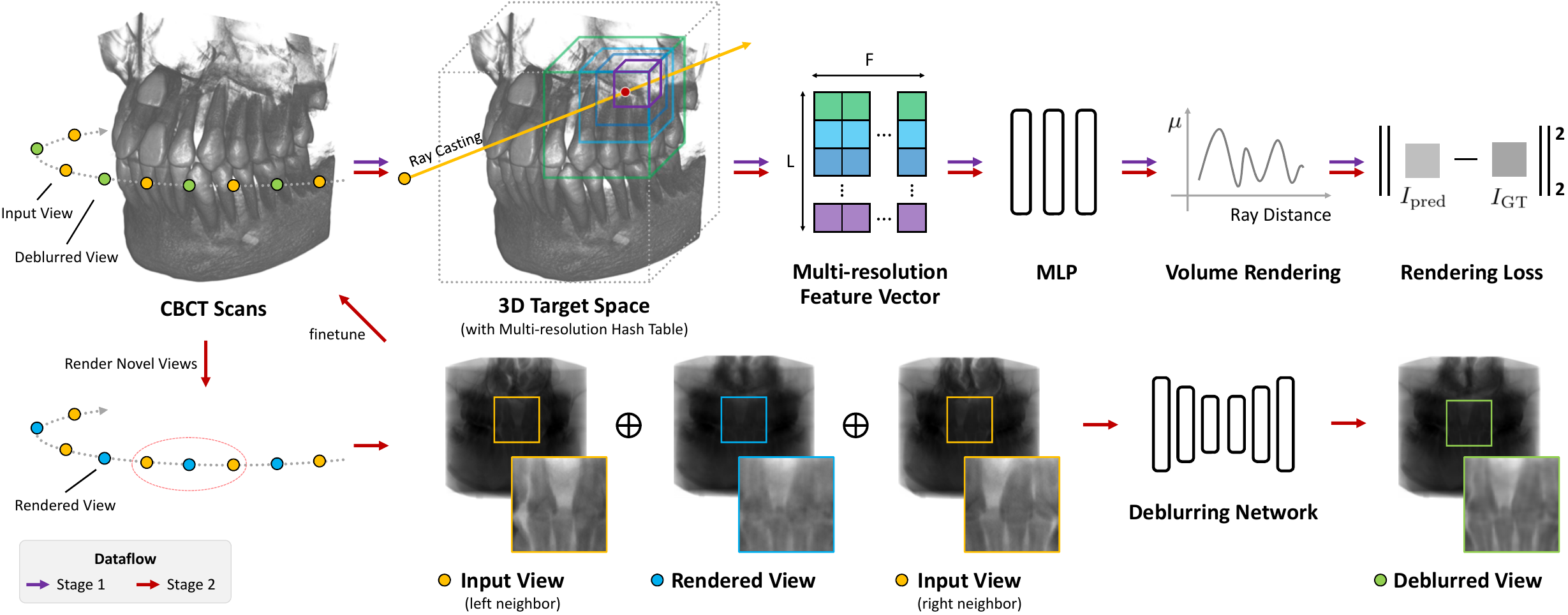}
   \caption{\textbf{Overview of our proposed {\em SNAF} for sparse-view CBCT reconstruction.} 
   Given the input views (orange dots), we first train an attenuation field with our equal-step ray casting (Sec.~\ref{sec:atte_field}) in stage one (purple arrows), such that we can generate several novel views (blue dots) in-between any pair of input. Then, in stage two (red arrows), we deblur rendered views using a newly proposed deblurring network (Sec.~\ref{sec:view_aug}). In return, the enhanced novel views (green dots) along with the input views serve as input to further fine-tune the learned attenuation field.
   }
   \label{fig:framework}
\end{figure*}

\section{Method}

Figure~\ref{fig:framework} illustrates an overview of our two-stage framework, where we first train an attenuation field in stage one, then fine-tune the trained field with the help of augmented novel views in stage two. We elaborate on the technical details in the following.

\subsection{Preliminary}
\label{sec:preliminary}
\textbf{CBCT Scanning.}
As illustrated in Fig.~\ref{fig:teaser}, the configuration of CBCT scanning is quite standard, where an X-ray source moves around a rotation axis along a planar arc trajectory to capture the 2D projections of the interested human organ (e.g., teeth). Specifically, the X-ray source emits a cone-shaped beam towards the interested region, and the isocenter is set as the rotation center which is roughly placed at the center of the interested region. When the scanning starts, the X-ray source moves from the beginning of the planned arc trajectory to the end with a fixed angular marching step, and the trajectory plane is perpendicular to the rotation axis. For example, it moves within the angular range [0, 210) with an angular step $10.5^\circ$, such that it will capture 20 2D scans. Note here, this is the most commonly used scanning configuration in clinical practice, denoted as \textbf{20-view}, other configurations, e.g., 30-view or 50-view, can be naturally derived with different angular marching steps. The angular range [0, 210) is adopted throughout this paper unless otherwise specified. Each 2D scan is a perspective projection from the emitted cone beam and is captured by a flat-panel detector on the opposite side reflecting the transmitted intensities through the interested region.

\textbf{CBCT and Attenuation Fields.}
In CBCT imaging, attenuation coefficient describes the fraction of X-rays being absorbed through different materials, dependent on their densities.
Attenuation field, akin to the radiance field, is the main focus of this paper.
Formally, our goal is to learn the mapping $(x, y, z) \to \mu$, that is, from a 3D point to its attenuation coefficient.

\subsection{Neural Attenuation Fields Learning}
\label{sec:atte_field}
In stage one, inspired by the vanilla NeRFs \cite{mildenhall2020nerf} and InstantNGP \cite{mueller2022instant}, we integrate the multi-resolution hash table into the conventional NeRF pipeline with some special adaptations to fit our unique setting, and leave the other core elements, e.g., MLPs unchanged.
Next, we introduce all the key adaptions.

\textbf{Uniform Sampling in Ray Casting.}
As has been mentioned in the introduction, a particular characteristic of true volume rendering is that there are no specific objects compared with the approximated surface rendering. Instead, the whole space of the interested region is informative. Thus, the two-pass importance sampling from NeRFs are not applied in our case, and we hope to cover the whole space using all sampling points along all rays. This implies that the sampling points along each ray are placed to be as even as possible. To this end, we design the equal-step ray casting in this way.

Given all the input 2D scans with corresponding camera parameters, we first initialize a discrete target volume (see Sec~\ref{sec:dataset} for detailed scales) accordingly as the space of the attenuation field. Then for each pixel in any scan, we shoot a ray $r(t) = o + td$ to the X-ray source, such that the ray intersects the bounding box of the volume twice, i.e., the incident point $o + t_{\min} d$ and exit point $o + t_{\max} d$. Finally, we sample points between $t_{\min}$ and $t_{\max}$ with a marching step $D$. It is clear that, different rays may obtain different numbers of inner sampling points and $D$ is set proportionally to the desired volume spacing.
Please refer to Sec.~\ref{sec:sampling} for our exhausted evaluation and discussion on different $D$ values and sampling strategies.

\textbf{Volume Rendering.}
For each 2D scan, the X-ray source emits a cone-shaped beam with an initial intensity, which is attenuated through different tissues encountered along each ray, and finally received by a flat-panel detector to form all the pixel-wise intensity values.
This rendering process is usually formulated by the Beer-Lambert Law:
\begin{equation}
\label{equ:beer}
    I = I_0 \ e^{- \sum{\mu_i \delta_i}},
\end{equation}
where $I$ is the transmitted intensity that can be read from the 2D scan, $I_0$ is the incident intensity, $\mu$ is the attenuation coefficient, while $\delta$ denotes the sampling marching step.

In this paper, we utilize Eq.~\ref{equ:beer} to produce the rendered (i.e., transmitted) image $I_{\text{pred}}$ from the attenuation field, and finally learn the neural attenuation field by minimizing the mean squared error:
\begin{equation}
    L = \sum_{r \in R_i} || I_{\text{pred}}(r) - I_{\text{GT}}(r) ||^2,
\end{equation}
where $R_i$ and $I_{\text{GT}}$ denote the set of rays and ground truth intensities, respectively.

Overall, we follow a similar training strategy to learn the attenuation field. Because of hash encoding, we can rapidly acquire a learned attenuation field with decent quality, and eventually get the reconstructed 3D volume by uniform grid sampling.

\subsection{Novel View Augmentation}
\label{sec:view_aug}
\begin{figure}[t]
  \centering
   \includegraphics[width=1.0\linewidth]{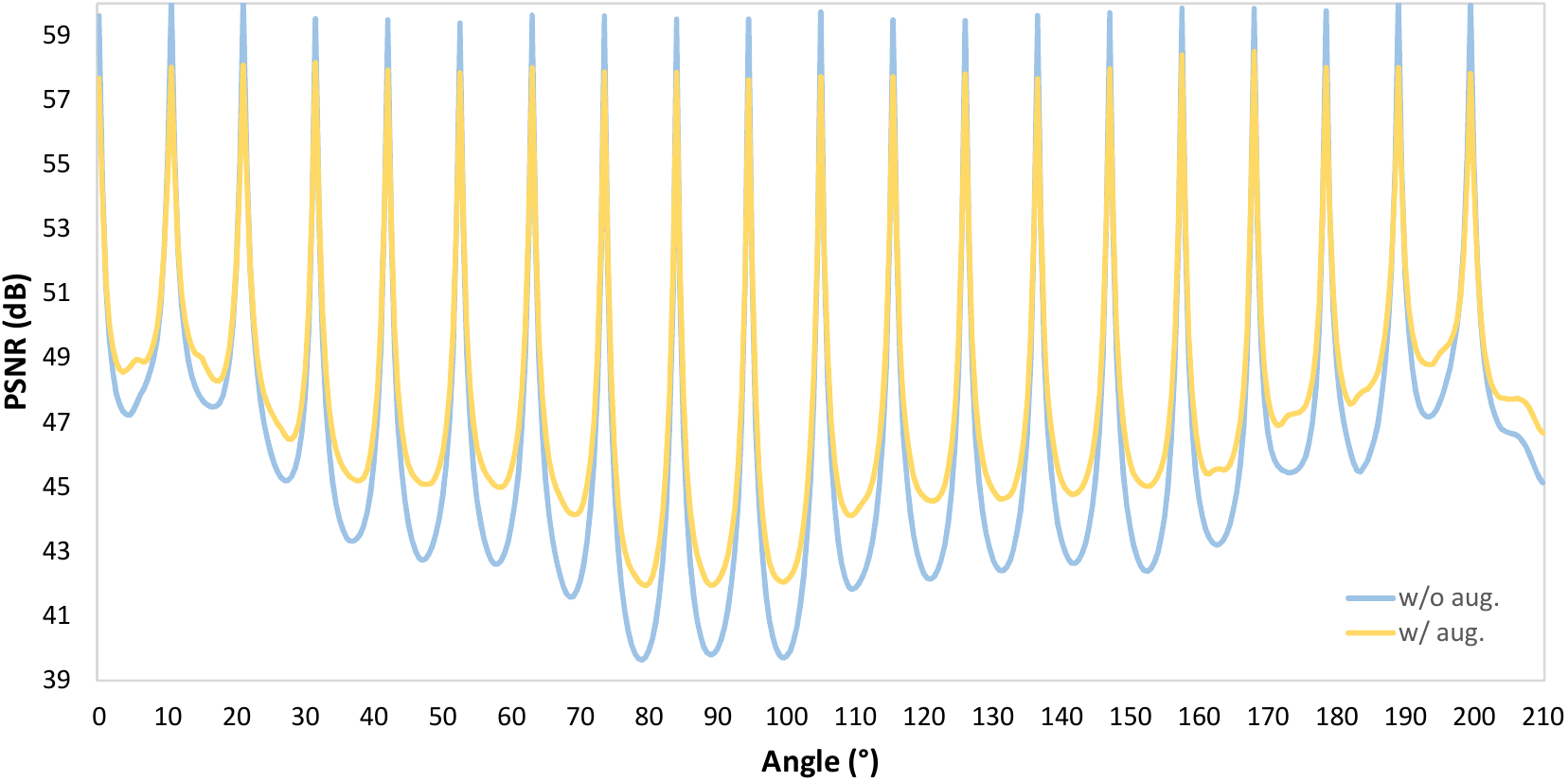}
    \caption{\textbf{Novel view prediction with/without view augmentation.} We visualize the generated novel views along the capturing trajectory in terms of the PSNR value. It is clear that there is significant degradation between the input and new views. And our view augmentation strategy remarkably reduces the gap. Note here, deblurring happens at the valley. }
   \label{fig:views}
\end{figure}

Even though the learned attenuation field in stage one can produce the CBCT image with decent quality, it suffers from severe blur caused by insufficient data from limited input views, see Fig.~\ref{fig:deblur} for an example. In the meanwhile, the existing techniques proposed for solving sparse-view challenges in neural rendering are not applicable as discussed in Sec.~\ref{sec:rw_sparseview}, we thus designed our novel view augmentation scheme based on our observations as introduced in the CBCT scanning process (Sec.~\ref{sec:preliminary}).

Specifically, we first render a set of novel views along the scanning trajectory, and quantize the image quality in terms of PSNR as shown in Fig.~\ref{fig:views}, the purple line, where we find that the quality degrades dramatically on the new views in-between any consecutive input views (see the valley) caused by blur (Fig.~\ref{fig:deblur}). Thus, a simple and effective idea is to deblur the in-between new views so as to improve the quality of the target attenuation field.

\textbf{Deblurring Network.}
Thanks to the unique configuration of the CBCT scanning, the neighboring views have strong appearance similarity and contain rich sharp features. To directly employ the image priors from the input views and maintain the geometry consistency of the interested region, we thus design a deblurring network that takes as input a rendered new view from the learned neural attenuation field in stage one, along with its left and right neighboring views via image concatenation. The goal is to map the blurry view to its high-quality counterpart with fine details.

\textbf{Data Preparation and Training.}
To train the deblurring network, we constructed a paired quadruple dataset from different patients. 
Specifically, we have collected data from 30 patients, for each patient's data, we train an attenuation field as described in stage one. We then produce an in-between new view $I_{\text{new}}$ for any neighboring input views $I_{\text{left}}$ and $I_{\text{right}}$. The corresponding ground truth views $I_{\text{gt}}$, along with the aforementioned three views form a paired quadruple $\{I_{\text{left}}, I_{\text{new}}, I_{\text{right}}; I_{\text{gt}}\}$ that is used to train the network. The network has a UNet structure~\cite{ronneberger2015u} with additional attention mechanism~\cite{vaswani2017attention, katharopoulos2020transformers}, and is pre-trained ahead. For detailed network structure, please refer to {\em supplementary materials}.

Having the pre-trained deblurring network, in stage two, we first deblur the new views, then the deblurred views as well as the original input views are sent to the neural network to fine-tune the learned attenuation field.
Note here, the generalization ability comes from the fact that for different patients, the CBCT scanning process will produce CBCT images with similar spatial and intensity ranges for the common interested region. We have validated the strategy thoroughly in Sec.~\ref{sec:results}.

\begin{figure*}[t]
  \centering
   \includegraphics[width=1.0\linewidth]{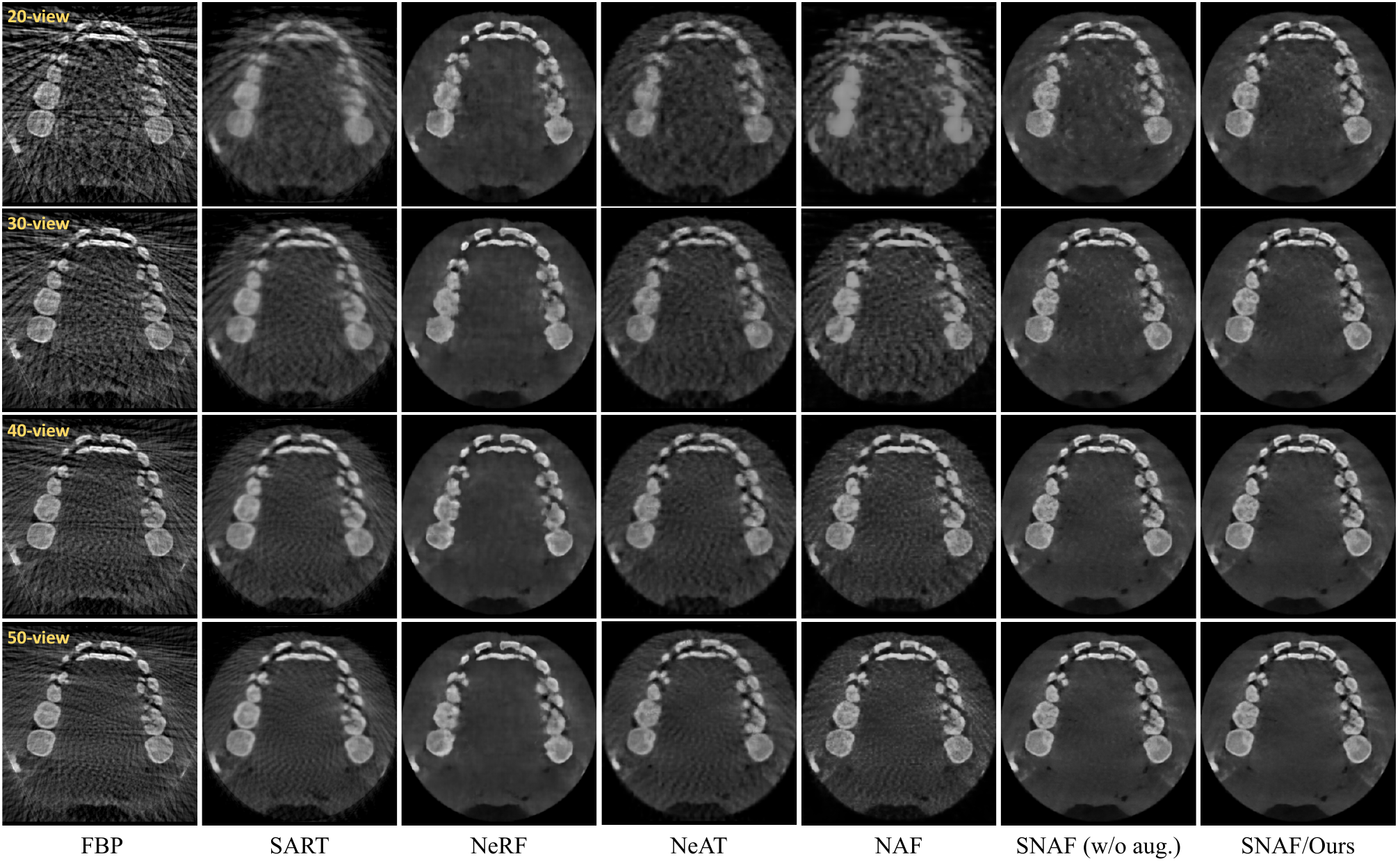}
   \caption{\textbf{Qualitative comparison with different methods.} We compare the performance of CBCT reconstruction with different input views (each row), and visualize slices of the reconstructed 3D volumes from different methods (each column). Our reconstructed quality dramatically outperforms all other methods. For reference, please see the ground truth in Fig.~\ref{fig:teaser} (axial slice). }
   \label{fig:comparison}
\end{figure*}

\begin{table*}[t]
  \centering
  \begin{tabular}{c|c c|c c|c c|c c}
    \toprule
    \multirow{2}{*}{Method} & \multicolumn{2}{c|}{20-view}  & \multicolumn{2}{c|}{30-view} & \multicolumn{2}{c|}{40-view} & \multicolumn{2}{c}{50-view}\\
    & PSNR $\uparrow$ & SSIM $\uparrow$ & PSNR $\uparrow$ & SSIM $\uparrow$ & PSNR $\uparrow$ & SSIM $\uparrow$ & PSNR $\uparrow$ & SSIM $\uparrow$ \\
    \midrule
    FBP~\cite{feldkamp1984practical}
    & 19.63 & 0.457 & 20.69 & 0.525 & 21.42 & 0.579 & 21.78 & 0.611 \\
    SART~\cite{andersen1984simultaneous} 
    & 26.29 & 0.808 & 27.29 & 0.835 & 28.00 & 0.856 & 28.43 & 0.870 \\
    NeRF~\cite{mildenhall2020nerf}
    & 30.02 & 0.910 & 31.26 & 0.929 & 31.77 & 0.936 & 31.98 & 0.938 \\
    NeAT~\cite{ruckert2022neat}
    & 26.40 & 0.863 & 27.65 & 0.882 & 28.29 & 0.890 & 28.34 & 0.891 \\
    NAF~\cite{zha2022naf}
    & 26.83 & 0.835 & 28.41 & 0.848 & 30.14 & 0.895 & 30.76 & 0.900 \\
    \midrule
    SNAF (w/o aug.)
    & 30.86 & 0.919 & 33.00 & 0.943 & 34.46 & 0.955 & 35.23 & 0.961 \\
    SNAF/Ours
    & \textbf{31.93} & \textbf{0.931} & \textbf{33.97} & \textbf{0.951} & \textbf{35.24} & \textbf{0.962} & \textbf{35.92} & \textbf{0.966} \\
    \bottomrule
  \end{tabular}
  \caption{\textbf{Quantitative comparison with different methods.} 
  We quantize the quality of the recosntructed 3D CBCT volumes from different methods using PSNR and SSIM metrics with different view configurations.
  Our method achieves the best results in both metrics for all configurations. Especially, our method with only 20 views outperforms all the others with 50 views.}
  \label{tab:comparison}
\end{table*}

\section{Results}
\subsection{Dataset}
\label{sec:dataset}
In our experiments, we adopt the dental dataset to validate the advantage of our proposed framework. 
For our dataset, the resolution of the 3D CBCT image is $401 \times 401 \times 401$ with a voxel-wise spacing of $0.2 \times 0.2 \times 0.2$ mm, and our synthetic 2D projections are of $404 \times 404$ with a spacing of $0.278 \times 0.278$ mm.
To synthesize the 2D projections, we apply volume rendering as per Eq.~\ref{equ:beer} with $I_0=1.0$ and simulate the scanning process with different angular ranges and angular marching steps.
And, for each ray in the ray casting, a fixed $D=0.1$mm (i.e., half of the CBCT spacing) is adopted, following the trilinear interpolation to calculate the attenuation value of each sampling point. 

\subsection{Network Implementation and Training}
In our framework, the multi-resolution hash table encoder and the MLPs are two crucial modules for learning neural attenuation fields.
For the hash table encoder, we adopt the official implementation of~\cite{mueller2022instant}, where set the core parameters to specific values to fit our setting, including the size $T=2^{19}$, the number of resolution levels $L=12$, and the feature dimension $F=8$.
By experiments, we observe that $F$ plays a more essential role in learning attenuation fields. Because, unlike the neural radiance fields in natural scenes that mainly representing object surfaces, our neural attenuation field is more complex and is informative within the entire scene with many anatomical details.
Consequently, neural attenuation fields require features with a larger size, and $F=8$ is used in all our experiments. For the fully connected layers, we use a compact 3-layer 64-channel MLP with ReLU to learn multi-resolution features from hash tables.

\textbf{Training.}
For training networks, Adam optimizer is used with an initial learning rate of 1e-3 and is decayed exponentially. We first pre-train the deblurring network with four Nvidia A100 GPU cards for 4 hours. Then, we train with one Nvidia A100 GPU card for 20 mins in stage one and 10 mins in stage two to learn and fine-tune the attenuation field.

\subsection{Comparison}
\label{sec:results}

\begin{figure}[t]
  \centering
   \includegraphics[width=1.0\linewidth]{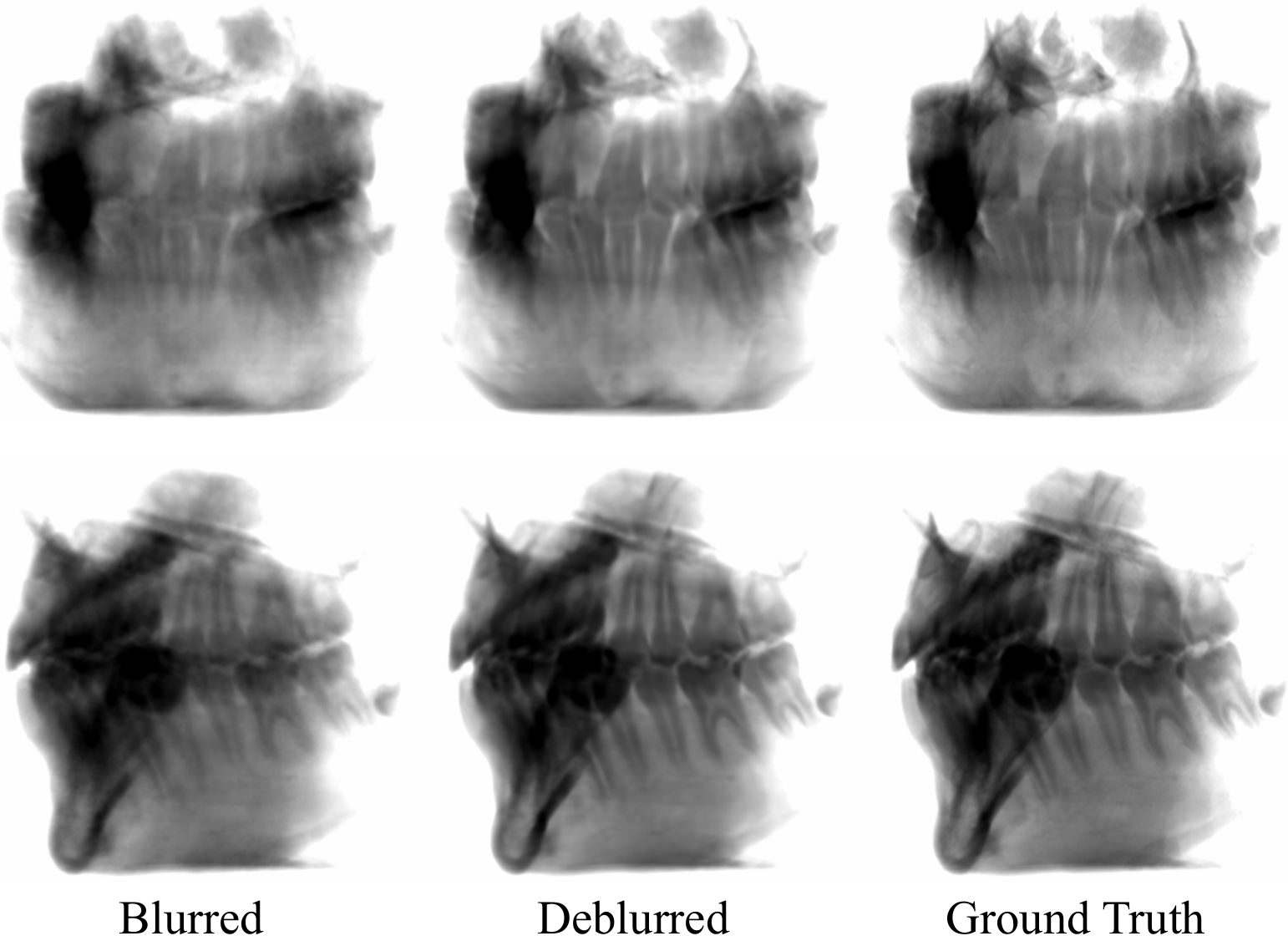}
   \caption{\textbf{Ablation studies on deblurring.} We visualize two typical examples of 2D views before (direct rendering, 1st column) and after deblurring (2nd column). The deblurring network greatly enhances sharp boundaries and details that comparable to the ground truth (3rd column).}
   \label{fig:deblur}
\end{figure}

To validate the advantage of our method,  we compare the our {\em SNAF} with both learning and non-learning methods, including Filtered Backprojection (FBP) ~\cite{feldkamp1984practical}, simultaneous Algebraic Reconstruction (SART) ~\cite{andersen1984simultaneous}, and three recently popular neural rendering solutions~\cite{mildenhall2020nerf, ruckert2022neat, zha2022naf}.
Specifically, FBP and SART are two traditional methods that are still commonly used in commercial CBCT systems, where FBP has high efficiency and SART is mainly designed for sparse-view CBCT reconstruction.
The neural rendering approach, Neural Radiance Fields (NeRF) \cite{mildenhall2020nerf}, is mainly designed for novel view synthesis in natural scenes, and we slightly modified it by removing its color prediction branch.
Furthermore, we also compare two concurrent neural volume rendering methods, Neural Adaptive Tomography (NeAT) \cite{ruckert2022neat} and Neural Attenuation Field (NAF) \cite{zha2022naf}, which are specifically designed for CBCT reconstruction with state-of-the-art performance.
For ablation purpose, we present a variant of our {\em SNAF} by removing the view augmentation, denoted as SNAF (w/o aug.).
The experiments are conducted in different numbers of input views (i.e., 20, 30, 40, and 50 views). Note that, 20-view is extremely sparse, and usually 25 times fewer than clinical collections.

Qualitatively, typical examples are displayed in Fig.~\ref{fig:comparison}, where it can be easily observed that our proposed method outperforms all competing methods, reconstructing CBCT images with more details and less noise.
Specifically, since FBP only performs well with sufficient views (e.g., 500 views in commercial CBCT systems), it produces many artifacts with 50 input views, and drops significantly when the view number decreases.
Another traditional method, SART, tackles the sparse view problem to a certain degree; however, it exceedingly trades reconstruction details with an over-smoothness problem that many tooth details, especially on 20 input views, are lost.
As for the three neural rendering methods (i.e., NeRF, NeAT, and NAF), NeRF generates fewer artifacts compared to its variants (i.e., NeAT and NAF) that are specifically designed for CBCT reconstruction. 
Unfortunately, the local intensity contrast (e.g., inside the tooth) cannot be reconstructed accurately due to its global optimization mechanism (i.e., without local feature embedding).
By comparison, SNAF (w/o aug.) already surpasses all other methods, providing high-quality visual results, while SNAF (ours) with novel view augmentation further improves the quality with more fine details. A separate visual example is shown in Fig.~\ref{fig:deblur} to better illustrate the difference, which in turn validates the effectiveness of our novel view augmentation strategy.

Quantitatively, PSNR and SSIM metrics are presented in Tab.~\ref{tab:comparison} to measure the reconstruction quality.
Compared to these methods, our framework achieves the leading performance with different numbers of input views.
An interesting observation is that SNAF (w/o aug.), running on 20 input views, even outperforms other approaches with 50 input views, except for NeRF (31.98 dB of PSNR).
Moreover, benefiting from the novel view argumentation, the reconstruction accuracy of ours is consistently improved in our sparse view settings (1.07 dB and 0.69 dB PSNR improvements of 20 views and 50 views, respectively), demonstrating the deblurring network can further enhance novel views, which in return fine-tune the learned attenuation field.

\textbf{Running Time.}
FBP is an efficient method that can reconstruct the 3D CBCT volume in 1 minute.
SART usually takes about 4-12 minutes for input views from 20 to 50.
While, for the neural rendering-based methods, NeRF is a global optimization with more training time (i.e., about 90 minutes using our dataset), and all its variants (NeAT, NAF, and our SNAF), are built on local feature encoding and take around 30 minutes on an A100 GPU.

\section{Ablation and Discussion}
We conduct extensive ablations and other experiments to discuss technical details (i.e., the point sampling strategy), the performance of our method under different scanning conditions (i.e., the scanning range), the performance with challenging metal artifacts, as well as the downstream dental applications. In this section, our default input view number is 20. Please refer to {\em supplementary materials} for more discussions.

\begin{table}[t]
  \centering
  \small{
    \begin{tabular}{c|c|c c||c c}
    \toprule
    \multirow{2}{*}{\makecell[c]{Interval \\ (mm)}} & \multirow{2}{*}{\makecell[c]{Time \\ (m:s)}} & \multicolumn{2}{c||}{Uniform} & \multicolumn{2}{c}{Stratified} \\
    & & PSNR $\uparrow$ & SSIM $\uparrow$ & PSNR $\uparrow$ & SSIM $\uparrow$ \\
    \midrule
    1.6 & 03:30 & 29.73 & 0.902 & 28.95 & 0.892\\
    0.8 & 05:30 & 30.63 & 0.917 & 30.13 & 0.912\\
    0.4 & 09:30 & 30.78 & 0.919 & 30.78 & 0.920\\
    0.2 & 17:30 & 30.86 & 0.919 & 30.88 & 0.920\\
    \bottomrule
  \end{tabular}
  }
  \caption{\textbf{Imaging evaluation with various marching steps.} We vary the sampling marching step in our equal-step sampling scheme, under the scanning condition of 20-view and a 3D spacing of 0.2 mm. We report the imaging quality with PSNR and SSIM of the reconstructed volume.}
  \label{tab:sampling}
\end{table}

\begin{figure}[t]
  \centering
   \includegraphics[width=1.0\linewidth]{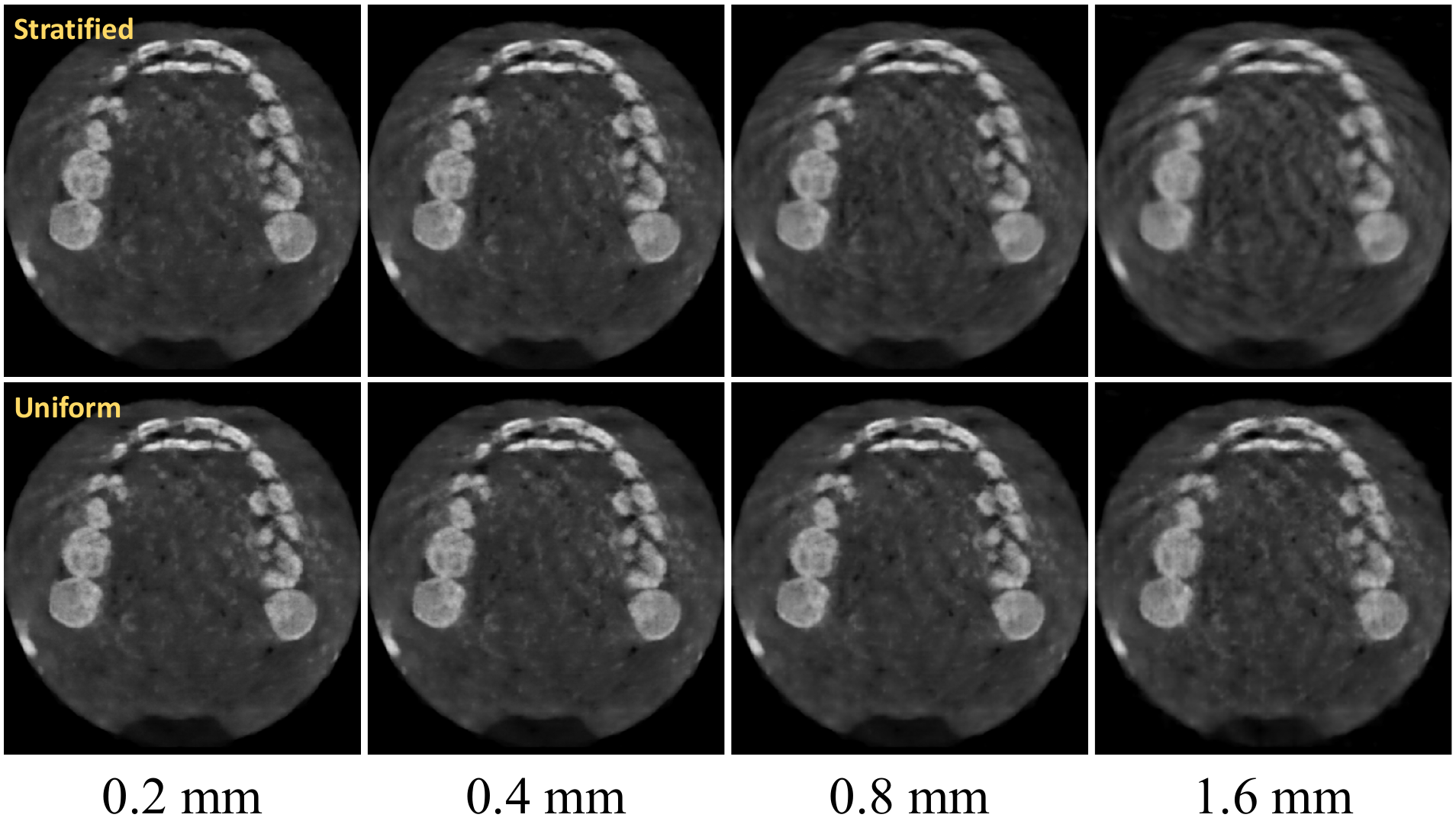}
   \caption{\textbf{Visual comparison of sampling strategies with various marching steps.}}
   \label{fig:sampling}
\end{figure}

\subsection{Point Sampling on Ray Casting}
\label{sec:sampling}
Sampling strategies are crucial in learning attenuation fields. In our {\em SNAF}, we adopt the simple uniform sampling strategy with a marching step $D$, although stratified sampling is an alternative solution. In this experiment, we validate our choice.

\textbf{Varying the marching step $D$}. 
It is clear that a bigger $D$ means fewer sampling points along a ray. With fewer points, it tends not to cover the entire space very well, but significantly simplifies the calculation. As can be seen from Tab.~\ref{tab:sampling}, when we increase $D$ from $0.2 \text{mm}$ to $0.8 \text{mm}$, marginal performance degradation is observed (30.86 vs. 30.63 for PSNR), while only less than 1/3 of the training time is required. Visual examples in Fig.~\ref{fig:sampling} also validates this point. Thus, to trade off the training time and reconstruction quality, we can properly and faithfully enlarge the marching step $D$, and $D=0.2 \text{mm}$ is used in our algorithm to first ensure reconstruction quality.

\textbf{Uniform vs. stratified sampling.}
We validate that the uniform sampling is the best fit in our attenuation field learning by comparing with the alternative stratified sampling scheme with increasing $D$ values. As can be seen from Tab.~\ref{tab:sampling} and Fig.~\ref{fig:sampling}, when we use smaller $D$, uniform sampling is on par with stratified sampling. However, large $D$, e.g., $1.6 \text{mm}$, will bring noticeable artifacts and significant decreasing in terms of PSNR (i.e., 29.73 vs. 28.95). The reasons are that, first, our final goal is to acquire a discrete representation (i.e., 3D volume) by grid sampling, rather than a completely continuous radiance field for generating novel views. 
It is reasonable to keep a fixed sampling step to capture this explicit 3D volume.
Second, uniform sampling allows integration of total variation loss along the ray, avoiding the streaking artifacts. Instead, the randomness introduced by the stratified sampling cannot produce sample points that are evenly distributed within the volume.
We thus conclude that uniform sampling is the best sampling strategy in learning neural attenuation fields.

\begin{table}[t]
  \centering
  \resizebox{\linewidth}{!}{
    \begin{tabular}{c|c c c c c c c}
    \toprule
    Range ($^\circ$) & 90 & 120 & 150 & 180 & 210 & 360\\
    \midrule
    PSNR $\uparrow$ & 27.13 & 28.76 & 30.36 & 30.73 & 30.86 & 30.95\\
    SSIM $\uparrow$ & 0.872 & 0.900 & 0.916 & 0.918 & 0.919 & 0.921\\
    \bottomrule
  \end{tabular}
  }
  \caption{\textbf{Imaging evaluation with various ranges.} We sample 20 views from increasing ranges of the scanning trajectory, and report PSNR and SSIM metrics of the reconstructed CBCT volume.}
  \label{tab:range}
\end{table}

\subsection{Scanning Range}
Scanning range that is measured by the rotation angle on the arc-shape trajectory, can vary flexibly, e.g., $180^{\circ}$, $210^{\circ}$(short scan), or $360^{\circ}$ (full scan). Given the same radiation dose (i.e., the same number of scanned views), bigger range can increase the view coverage, while significantly takes more scanning time. In this experiment, we intend to determine the best scanning range in our sparse-view setting to trade off scanning time and reconstruction quality. As can be seen from Tab.~\ref{tab:range}, scanning ranges smaller than $150^{\circ}$ will lead to obvious performance degradation, while from $180^{\circ}$ to $360^{\circ}$, the improvement is subtle. We thus recommend $210^{\circ}$ as our best scanning range in all our experiments, which is also consistent with the choice in clinical practice when capturing dense views.

\subsection{Experiments on Metal Artifacts}
\begin{figure}[t]
  \centering
   \includegraphics[width=1.0\linewidth]{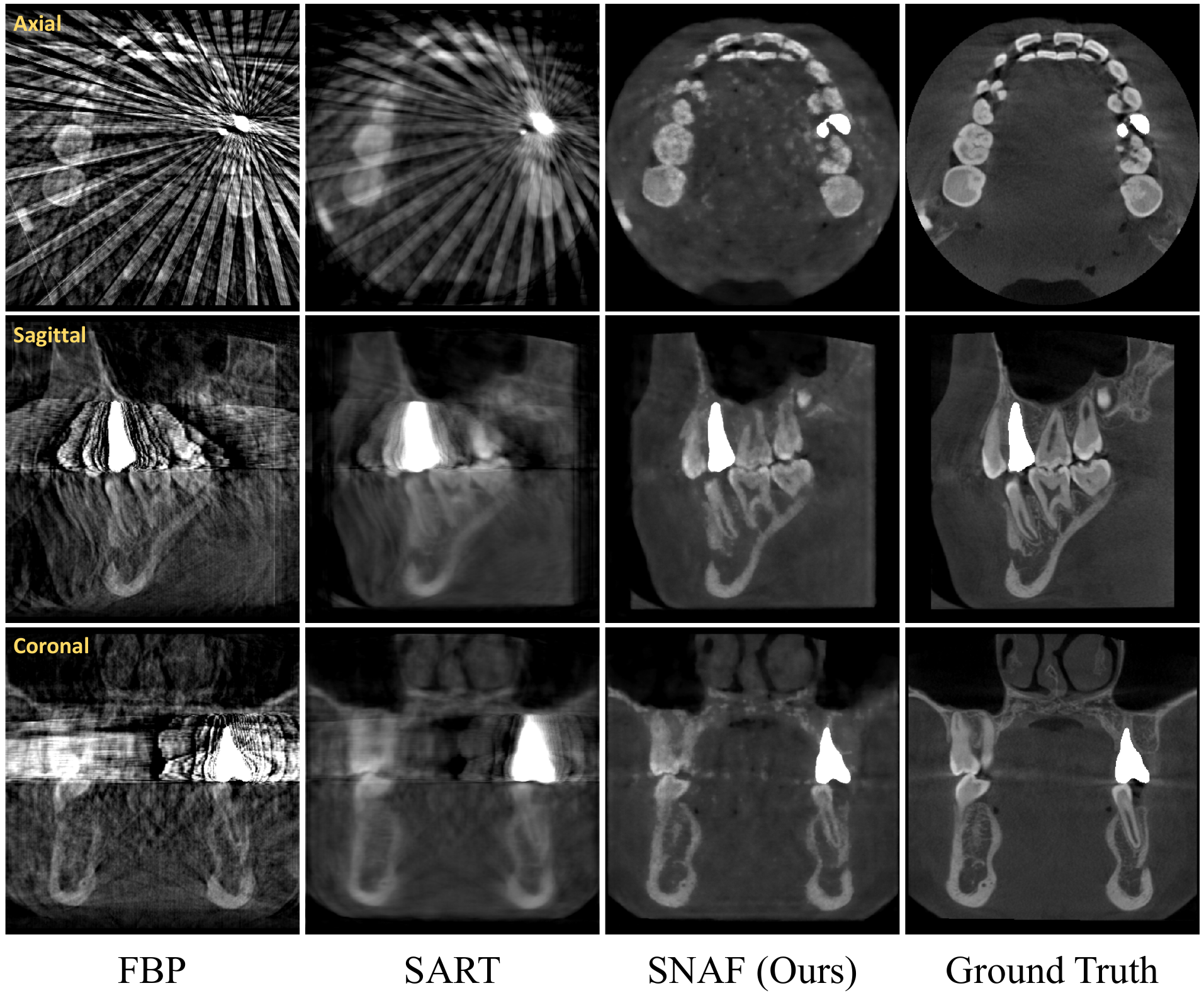}
   \caption{\textbf{Visual results with metal artifacts.} 
    We synthesize input views with metal by putting a synthetic metal into a clean ground truth (last column) CBCT volume. Given such input, our SNAF can faithfully eliminate almost all metal artifacts, while alternative methods (i.e., FBP and SART at the first and second columns) suffer from severe streaking artifacts.
    }
   \label{fig:metal}
\end{figure}

Metal artifact is a long-standing problem in CBCT, causing by the metallic materials, e.g., metal implants. In this experiment, we demonstrate that the learned neural attenuation fields can significantly eliminate metal artifacts.
Specifically, we first select a clean 3D CBCT volume without any artifacts, then simulate a metal implant by turning a tooth into metal, i.e., assigning a large attenuation value, as shown in the last column of Fig.~\ref{fig:metal}.
We use the 20-view configuration to generate 20 views of this synthetic CBCT volume, and compare the reconstructed results of FBP, SART, and ours.

In Fig.~\ref{fig:metal}, we find that FBP produces bright streaking artifacts and reconstructs unusable results.
Although, SART manages to ease these artifacts by smoothness, still cannot hinder the degradation.
By comparison, our method with learned neural attenuation fields reconstructs CBCT images with high-quality, where the boundaries near metal area are even clear.
The main reason is that other methods force to accumulate the transmitted intensities by backprojection from given views, polluting the reconstruction with affected transmitted intensities.
Instead, neural attenuation fields can faithfully compute attenuation coefficients for any sampling point, no matter whether they are metal or not.

\subsection{Applications}
\begin{figure}[t]
  \centering
   \includegraphics[width=1.0\linewidth]{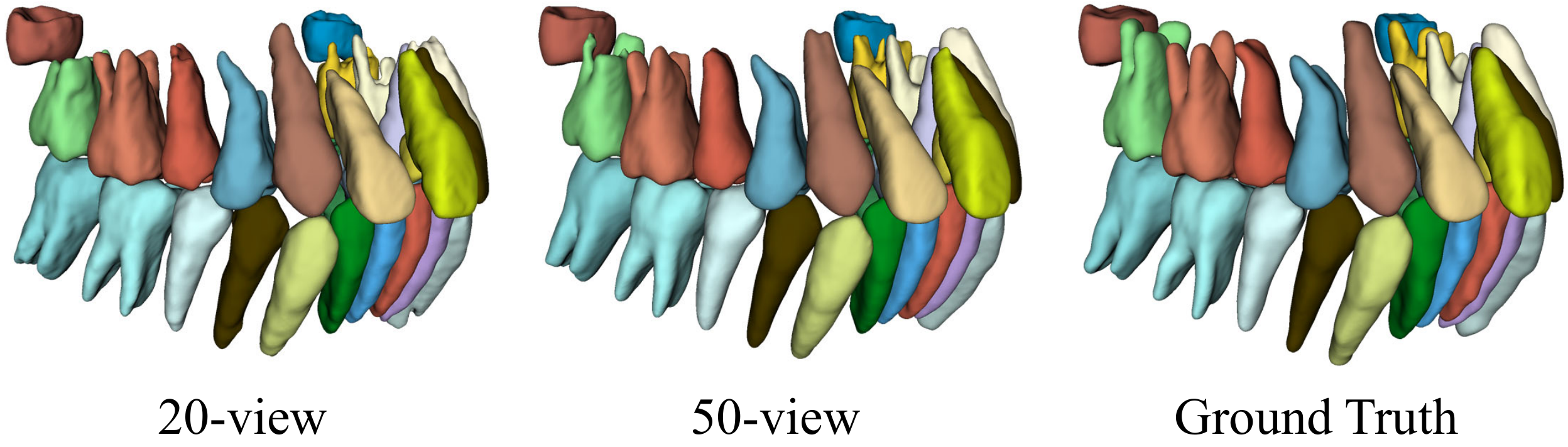}
   \caption{\textbf{Visual results of tooth instance segmentation on reconstructed volumes from SNAF.}
   }
   \label{fig:seg}
\end{figure}

Our {\em SNAF} targets the high-quality CBCT reconstruction with sparse input views. In order to validate that our result can be used in real clinics, we perform tooth semantic segmentation~\cite{cui2022fully} directly on the reconstructed CBCT image. Example segmentation results are shown in Fig.~\ref{fig:seg}, where all teeth are successfully from the 20-view and 50-view results. Even with only 20 views, most of the teeth can be recognized and segmented faithfully, although the roots of the molar teeth are imperfect because of the blurry signal in the reconstructed CBCT image.

\subsection{Limitations and Future Work}
Although SNAF provides high-quality CBCT reconstruction with only 20 views, it can fail when input data is extremely insufficient (e.g., 10 views).
We plan to integrate generative networks with neural attenuation fields to tackle this big challenge in future works.
Besides, we produce one novel view in-between any consecutive input views, instead of adaptively selecting novel views in terms of reconstruction quality. An observation from Fig.~\ref{fig:views} is that the valley of the PSNR distribution corresponds to a Gaussian distribution, since front views contain more information than side views and more views should be placed around those positions. Luckily, the Gaussian distribution can serve as a density function to guide the advanced adaptive novel view generation in the future. 

\section{Conclusion}
In this paper, we present {\em SNAF}, a novel approach for CBCT reconstruction from sparse views by neural attenuation field learning. A new view augmentation scheme is proposed to deal with the great challenges introduced by insufficient data. Extensive results and experiments validate the high-quality of our results.
We believe that neural attenuation fields are promising to benefit other medical image processing tasks, providing its potential in future CBCT scanning system.

%%%%%%%%% REFERENCES
{\small
\bibliographystyle{ieee_fullname}
\bibliography{egbib}
}

\end{document}